\documentclass[10pt]{article}
\usepackage{anysize}
\marginsize{1cm}{1cm}{2.5cm}{2.5cm}
\usepackage{amsfonts}
\usepackage{multicol}
\usepackage{graphicx} 
\DeclareGraphicsExtensions{.bmp,.png,.pdf,.jpg}
\usepackage{epsfig}
\usepackage{latexsym}
\usepackage{amsmath, amsthm, amsfonts, amssymb}
\usepackage{verbatim}
\usepackage{appendix}
\usepackage{epstopdf}
\usepackage{titlesec}
\usepackage{times}
\usepackage[Sonny]{fncychap}
\renewcommand{\thesection}{\Roman{section}}

\newtheorem{theorem}{Theorem}

\begin{document}

\pagenumbering{arabic}
\title{Mixed dynamics from the classical and quantum ergodic hierarchy}
\author{\textsc{Ignacio S. Gomez}$^{1}$ and \textsc{Federico H. Holik}$^{2}$}
\maketitle

\begin{abstract}
Based on the classical and quantum ergodic hierarchy, a framework for mixed systems with a phase space composed by two uncorrelated integrable and chaotic regions is presented. It provides some features of mixed systems connecting the intuitive notion of a mixed phase space with the mixing level of the ergodic hierarchy. The formalism is illustrated with the kicked rotator.\end{abstract}

\begin{center}
{\small 1- Departamento de Ciências Exatas e Naturais, Universidade Estadual do Sudoeste da Bahia,
			    BR 415, Itapetinga - BA, 45700-000, Brazil.\\[0pt] 2- Instituto de Física La Plata, CONICET - UNLP, 1900 La Plata, Argentina. \\[0pt]}

{\small
\centerline{\emph{Key words:
ergodic, mixing, EH, QEH, mixed system, quantum chaos}} }

\end{center}

\titleformat{\section}{\normalfont\scshape}{\thesection}{1em}{}

\maketitle


\section{Introduction}

Quantum chaos is well-known as the study of the quantum mechanical aspects of quantum systems which have a chaotic classical description \cite{stockmann,haake,gutzwiller,casati,tabor}. There are several approaches which give a chaotic classical description and these are related to each other: the algorithmic
complexity \cite{BELOT}, the Liapunov exponents \cite{lich} and
the Ergodic Hierarchy \cite{3M}. The Ergodic Hierarchy (EH) ranks the chaos of dynamical systems
through the cancellation of the correlation $C(A,B)$ between any two
subsets $A$ and $B$ of the phase space for large times $t\rightarrow\infty$, where $C(A,B)=\mu(A\cap B)-\mu(A)\mu(B)$. In \cite{NACHOSKY MARIO} we proposed a quantum generalization, named \emph{The Quantum Ergodic Hierarchy} (QEH) \cite{0, NACHOSKY MARIO} which expresses the cancellation of the correlations between states and observables for large times. The cancellation of these quantum correlations occurs when any quantum state $\rho$ reaches the relaxation and it admits a representation by an equilibrium state $\rho_{\ast}$ called \emph{weak limit}, with the relaxation expressed in terms of a relevant subalgebra $\mathcal{O}_R$ of the quantum algebra $\mathcal{A}$ \cite{vkampen, daneri, zeh, vhove, Omnes}. In \cite{NACHOSKY MARIO} we have described the chaotic behavior of the Casati-Prosen model \cite{casati verdadero, casati model} and the kicked rotator \cite{stockmann, haake} in terms of the QEH levels. Consequences of the QEH have been explored regarding graininess of phase space 
\cite{nacho1}, spectral decomposition theorem of dynamical systems \cite{nacho2}, gaussian ensembles \cite{nacho3}, KS-entropy \cite{nacho4}, quantum chaos timescales \cite{nacho5,nacho6} as well as connected developments involving the Pesin Theorem \cite{nacho7} and Lyapunov exponents in a non-Hermitian dynamics \cite{nacho8}. The kicked rotator presents general features of the mixed systems, i.e. the systems which have a phase space that has integrable and chaotic regions simultaneously. This is a typical situation in quantum chaos, where the system is neither integrated nor completely chaotic (see for example \cite{stockmann} pag. 92), and its nearest neighbor spacing distribution is between Poison and Wigner-like behavior. Related to this, there exist several phenomenological and theoretical approaches to these systems \cite{Bro73, Len91, Rob83, Ber84}. In this letter we study the mixed systems from the EH and QEH viewpoint, and we consider that the Hamiltonian is $H=H_0+\lambda V$, where $H_0$ is the unperturbed Hamiltonian, $\lambda V$ is a perturbation and $\lambda$ is a continuous parameter. In such case, chaotic transitions occur for a certain critical value $\lambda_c$ where the phase space goes from an integrable regime to a chaotic one (Fig. 1.3 of \cite{stockmann}). Complementarily, the geometric structure of the quantum space state is convex, thus manifesting the convexity is an important feature of quantum mechanics \cite{MielnikGQS, Holik}. Then, any geometric characterization of some aspect of quantum mechanics must be consistent with the convex structure of the space state. In particular, any geometrical characterization of quantum chaos should satisfy this requirement. In this letter, combining the formalism of the EH and QEH with the convexity of quantum space state we obtain a framework of mixed systems provided with a geometrical characterization. The work is organized as follows. In section 2 we present the formalism and we review the ergodic and mixing levels of the QEH. In section 3 we rewrite the ergodic and mixing levels of the QEH in terms of convex subsets and we present general features of mixed systems. In section 4 we apply the formalism to the kicked rotator. Finally, in section 5 we draw our conclusions.

\section{Preliminaries}
We begin introducing the formalism and mathematical tools used in this work. We use the formalism for states and observables established by the Brussels school (led by Ilya Prigogine) in \cite{bruselas}. The quantum state space is defined by the positive cone of functionals,
i.e. $\mathcal{C}=\{\rho \in \mathcal{A}^{\prime}:\rho\geq 0 \,\,,\rho(\mathbf{1})=1, \,\,
\rho^\dag=\rho\}$ where $\mathcal{A}$ is the quantum algebra of observables and $\mathcal{A}^{\prime}$ is the dual space of $\mathcal{A}$. 
We denote the states by $\rho$, the observables by $O$ and the mean value of an observable $O$ in a state $\rho$ at time $t$ (i.e. $\langle O\rangle_{\rho(t)}$) by $(\rho(t)|O)$. Considering the case
$O=\mathbf{1}$ we obtain the normalization condition for the states $Tr(\rho)
=(\rho|\mathbf{1})= \rho(\mathbf{1})=1$.

A way to obtain the equilibrium arrival of a quantum system is to cut the observable space $\mathcal{A}$ in order to obtain a relevant observable subalgebra $\mathcal{O}_R\subseteq \mathcal{A}$. Physically, $\mathcal{O}_R$ contains all the relevant information of the system to which we can have access. The subalgebra $\mathcal{O}_R$ can be chosen in different ways \cite{vkampen, daneri, zeh, vhove, Omnes}. In this context it is possible to define the \emph{relaxation} of a state $\rho$ as follows. We say that
$\rho$ reaches the relaxation state $\rho_R$ at $t=t_R$ ($t_R\in\mathrm{R}_{\geq0}$ or $t_R=\infty$) if and only if 
\begin{equation}\label{relaxation1}
(\rho(t)|O)=(\rho_{R}|O) \,\,\,\,\,\ \forall t\geq t_R, \,\,\ \forall O\in\mathcal{O}_R.
\end{equation}
The QEH expresses the relaxation in terms of the weak limit and the C\`{e}saro limit. Given a state $\rho$ we define the weak limit and
the C\`{e}saro limit as follows. We say that $\rho_{\ast}$ is the weak-limit($O$) of $\rho$
if, for any $O\in\mathcal{O}_{R}$, the following relation is
satisfied
\begin{equation}\label{weak limit-1}
\lim_{t \rightarrow\infty}(\rho(t)|O)=(\rho_{\ast}|O)
\end{equation}

\noindent The weak-limit(${\mathcal{O}_R}$) is symbolized as

\begin{equation}\label{weak limit-2}
W-\lim_{t\rightarrow\infty}\rho(t) = \rho_{\ast}
\end{equation}
\noindent According to this definition, the
weak limit is the relaxation state at $t_R=\infty$. That is, $\rho_{*}$ is the relaxation state for large
times. On the other hand, we say that $\rho_{C}$ is the
C\'{e}saro-limit(${\mathcal{O}_R}$) of $\rho$ if, for any $O\in
\mathcal{O}_R$, the following relation is satisfied
\begin{equation}\label{cesaro limit-1}
lim_{N \rightarrow\infty}\frac{1}{N}\sum_{j=0}^{N-1}(\rho(j)|O) =
(\rho_{C}|O)
\end{equation}
The C\'{e}saro-limit(${\mathcal{O}_R}$) is symbolized as
\begin{equation}\label{cesaro limit-2}
C-lim_{t \rightarrow\infty}\rho(t) = \rho_{C}
\end{equation}

\noindent We note if $\mathcal{O}_R=\mathcal{A}$ then the above
definitions coincide to those given in \cite{0, NACHOSKY MARIO}. The weak and C\`{e}saro
limits(${\mathcal{O}_R}$) are equilibrium states in the sense of the mean values. 

We review the Quantum Ergodic Hierarchy \cite{0, NACHOSKY MARIO} that defines
the different chaotic levels through the cancelation of the quantum correlations between states and observables for large times ($t\rightarrow\infty$). We only use the ergodic and mixing levels whose definitions
are given in terms of the weak and C\`{e}saro limits \cite{0, NACHOSKY MARIO}. Given a set of relevant observables $\mathcal{O}_R\subseteq
\mathcal{A}$ we extend the ergodic and mixing definitions (see sections 3 and 4 of \cite{NACHOSKY MARIO}) as follows. Let $S$ be a quantum system and let $\mathcal{C}$ be its space
state. We say that $S$ is \emph{ergodic}(${\mathcal{O}_R}$) if all state $\rho\in\mathcal{C}$ has a
C\'{e}saro-limit(${\mathcal{O}_R}$) $\rho_{*}\in\mathcal{C}$ (see equations \eqref{cesaro limit-1} and \eqref{cesaro limit-2}). On the other hand, we say that $S$ is \emph{mixing} (${\mathcal{O}_R}$) if all state $\rho\in\mathcal{C}$ has a weak limit(${\mathcal{O}_R}$) $\rho_{*}\in\mathcal{C}$ (see equations \eqref{weak limit-1} and \eqref{weak limit-2}). For a physical interpretation of the ergodic and mixing levels we define the correlation between a
state $\rho$ and an observable $O$ \footnote{We omit
the subscript $R$ in $\mathcal{O}_R$ and we mean that an observable $O$ is in relation to
a given subalgebra of relevant observables $\mathcal{O}_R$. We consider that
$\mathcal{O}_R$ is not trivial i.e. $\mathcal{O}_R\neq\{0,I\}$.}. We define the quantum correlation between
$\rho(t)$ and an observable $O$ in the sense of the quantum ergodic hierarchy as

\begin{equation}\label{qcorrelation}
C_Q(\rho(t),O)=(\rho(t)|O)-(\rho_{*}|O)
\end{equation}

\noindent From these definitions we can also define the ergodic
and mixing levels in terms of $C_Q$ 

\begin{equation}\label{qcorrelation2}
\begin{split}
&\textrm{S is ergodic} \ \Longleftrightarrow \,\ lim_{N\rightarrow
\infty}\frac{1}{N}\sum_{j=0}^{N-1}C_Q(\rho(j),O)=0\\
&\textrm{S is mixing} \ \Longleftrightarrow \,\ lim_{t\rightarrow
\infty}C_Q(\rho(t),O)=0
\end{split}
\end{equation}
\noindent The equation \eqref{qcorrelation2} says that for large times the correlations vanish in mixing systems while in ergodic systems these vanish on average.

\section{Mixed systems and geometrical characterization of the QEH}

We express the ergodic and mixing levels of the QEH in terms of operators whose domains are convex
subsets, thus allowing a characterization of the chaotic transitions through the geometry of quantum space state. We consider a quantum system $S$ with  Hamiltonian $H=H_0+\lambda V$, where $\lambda V$ is a perturbation and $\lambda$ a continuous parameter. We define the following maps. Let $\rho\in\mathcal{C}$ be a state and let $O\in\mathcal{O}_R$ be an
observable. Given $\lambda$, the operators
$F_{\infty,\lambda}^{(M)}:\mathcal{C}\rightarrow\mathcal{C}$ and
$F_{\infty,\lambda}^{(E)}:\mathcal{C}\rightarrow\mathcal{C}$ are defined by
\begin{equation}\label{convex limits1}
\begin{split}
&F_{\infty,\lambda}^{(E)}(\rho)=lim_{N\rightarrow\infty}\frac{1}{N}\sum_{i=0}^{N-1}(\rho_{\lambda}(i)|O)\\
&F_{\infty,\lambda}^{(M)}(\rho)=lim_{t\rightarrow\infty}(\rho_{\lambda}(t)|O)
\end{split}
\end{equation}
where $\rho_{\lambda}(t)=U_{\lambda}(t)\rho(U_{\lambda}(t))^{\dag}$
and $U_{\lambda}(t)$ is the evolution operator. The limits of \eqref{convex limits1} are linear and therefore, $F_{\infty,\lambda}^{(M)},F_{\infty,\lambda}^{(E)}$ are linear operators. Moreover,
$Dom(F_{\infty,\lambda}^{(M)})$ and $Dom(F_{\infty,\lambda}^{(E)})$
are convex subsets of the space state. 
From these observations and the definitions of the previous section we can rewrite the ergodic and
mixing levels of the Quantum Ergodic Hierarchy in terms of the
convex sets $\mathcal{C}_{\lambda}^{(E)}=\{\rho\in \mathcal{C}: \rho \ \textrm{has Césaro-limit}\}$ and
$\mathcal{C}_{\lambda}^{(M)}=\{\rho\in \mathcal{C}: \rho \ \textrm{has weak limit}\}$:

\begin{equation}\label{convex definitions}
\begin{split}
&\textrm{a quantum system S is \textbf{ergodic}} \ \Longleftrightarrow \,\ \mathcal{C}_{\lambda}^{(E)}=\mathcal{C}\\
&\textrm{a quantum system S is \textbf{mixing}} \ \Longleftrightarrow \,\
\mathcal{C}_{\lambda}^{(M)}=\mathcal{C}
\end{split}
\end{equation}

\noindent Therefore, $\mathcal{C}_{\lambda}^{(E)}$ and $\mathcal{C}_{\lambda}^{(M)}$ express the first two levels, ergodic and mixing, of
the QEH in a geometrical form. On the other hand, if we assume that $S$ has a classical limit $S_{classical}$ then we have that $S_{classical}$ is a dynamical system $(\Gamma, \Sigma, \mu, \{T_t\}_{t\in \mathbb{R}})$\footnote{Where the $\sigma$-algebra $\Sigma$ is the power set of $\Gamma$, i.e. $\Sigma=\mathcal{P}(\Gamma)$ and $T_t$ is the classical Liouville transformation.} where the phase space $\Gamma$ is typically composed by integrable regions and chaotic regions simultaneously (see \cite{stockmann} Fig.1). Given $\lambda$, we can reasonably assume that the chaotic and the integrable regions are uncorrelated \footnote{This assumption is used by Robnik and Berry in \cite{Ber84}.} and we consider that $S_{classical}$ can be modelled by a normalized measure
$\mu_{\lambda}=\mu_{0}+\mu_{M}$ concentrated in two sets $E_{\lambda}$ and $A_{\lambda}$. For all $\lambda$, $E_{\lambda}$ and $A_{\lambda}$ are the regular and chaotic regions respectively, and we define the \emph{conditions of an uncorrelated mixed phase space} given by 

\begin{equation}\label{mixed1}
\begin{split}
&\Gamma =E_{\lambda}\cup A_{\lambda} \,\,\,\,\,\,\,\,\,\,,\,\,\,\,\,\,\,\, E_{\lambda}\cap A_{\lambda}=\emptyset\\
&\mu(E_{\lambda})=\mu_{0}(E_{\lambda}) \,\,\,\,\,\,\,\,\,\,,\,\,\,\,\,\,\,\, \mu(A_{\lambda})=\mu_{M}(A_{\lambda})\\
&\mu_{0}(A_{\lambda})=\mu_{M}(E_{\lambda})=0\\
&\mu_{0}|_{E_{\lambda}} \quad \textrm{regular dynamics}, \quad \mu_{M}|_{A_{\lambda}} 
\quad \textrm{mixing dynamics}
\\
\end{split}
\end{equation}
This set of conditions models a chaotic system with a mixed phase space within the framework of the ergodic hierarchy. The first three equations say that the phase space is divided into two disjoint and uncorrelated regions and the last equation fixes the dynamical level corresponding to these regions. To complete the description of the mixed system, we define the \emph{general critical conditions} of the chaotic transitions when $\lambda$ continuously varies from $0$ to $\lambda\gg \lambda_c$. 
\begin{equation}\label{mixed2}
\begin{split}
&lim_{\lambda\rightarrow0}\mu(A_{\lambda})=0 \quad (\textrm{regular regime conditions  $\lambda\ll1$})\\
&lim_{\lambda\rightarrow0}\mu(E_{\lambda})=1\\
&\mu(A_{\lambda})\longrightarrow 1 \quad  \textrm{when} \quad \lambda\gg \lambda_c \quad  (\textrm{chaotic regime conditions \,\ $\lambda\gg \lambda_c$})\\
&\mu(E_{\lambda})\longrightarrow 0 \quad \textrm{when} \quad \lambda\gg \lambda_c\\
&\frac{\partial^{2} \mu (A_{\lambda})}{\partial \lambda^{2}}|_{\lambda=\lambda_c}=0 \quad  (\textrm{phase chaotic transition when $\lambda=\lambda_c$})
\end{split}
\end{equation}
The equation \eqref{mixed2} is an asymptotic characterization of the chaotic transitions where the last equation is inspired by the phase transitions of thermodynamics represented by the behavior of the second derivatives of the specific heat. Making an analogy, we postulate that the critical perturbation parameter $\lambda_c$ behaves in the same manner as the critical temperature  $T_c$ in phase transitions of statistical mechanics.   
The following two theorems characterize the dynamical of the chaotic transitions based on the EH and QEH framework.

\begin{theorem}(Chaotic transitions as a function of the volume of the chaotic region in the integrable regime $\lambda\leq\lambda_c$)\label{teo1} \\ If $\mu_{M}|_{A_{\lambda}}$ is mixing, $lim_{\lambda\rightarrow0}\mu(A_{\lambda})=0$ and $\frac{\partial^{2} \mu (A_{\lambda})}{\partial \lambda^{2}}|_{\lambda=\lambda_c}=0$ then
\begin{equation}\label{teo1-1}
\mu(A_{\lambda})\simeq\mu(A_{\lambda_c})\left(\frac{3}{2}\left(\frac{\lambda}{\lambda_c}\right)^{2}-\frac{1}{2}\left(\frac{\lambda}{\lambda_c}\right)^{3}\right) \quad \textrm{for} \quad  \lambda \in[0, \lambda_c+\varepsilon]
\end{equation}
where $\lambda_c$ is the critical value of the chaotic transition. In particular, for small perturbations $\lambda\ll1$ we have
\begin{equation}\label{teo1-2}
\begin{split}
\frac{vol(\mathcal{C}_{\lambda}^{(M)})}{vol(\mathcal{C})}\propto\mu(A_{\lambda})\simeq \mu(A_{\lambda_c})\frac{3}{2}\left(\frac{\lambda}{\lambda_c}\right)^{2} \quad \textrm{for} \quad \lambda\ll1 \\
\end{split}
\end{equation}
where $vol(.)$ is any notion of volume defined over the subsets of the space state.
\begin{proof}
It is clear that the first equation of \eqref{mixed2} implies $\mu(A_{\{\lambda=0\}})=0$ (absence of chaos when $\lambda=0$). Now we are going to show that
$\frac{\partial \mu (A_{\lambda})}{\partial \lambda}|_{\lambda=0}=0$. This is a consequence of the mixing level. Let $A_{\{\lambda_n=\frac{1}{n}\}}$ be a decreasing sequence of sets $A_1\supsetneq A_{\frac{1}{2}} \supsetneq...\supsetneq A_{\frac{1}{n}}...\Rightarrow A_{\frac{1}{n}}=A_1\cap A_{\frac{1}{2}} \cap...\cap A_{\frac{1}{n}}$ and where $\mu(A_1)<1$. Then we have $\frac{\mu (A_{\frac{1}{n}})-\mu(A_{\{\lambda=0\}})}{\frac{1}{n}-0}=n\mu(A_{\frac{1}{n}})=n\mu(T_t A_{\frac{1}{n}})=n\mu(T_t(A_1\cap A_{\frac{1}{2}} \cap...\cap A_{\frac{1}{n}}))=n\mu(T_t A_1\cap T_t A_{\frac{1}{2}} \cap...\cap T_t A_{\frac{1}{n}}))$ for all $t\in\mathbb{R}$ since $T_t$ is a bijective measure preserving transformation. In particular, we have $\frac{\mu (A_{\frac{1}{n}})-\mu(A_{\{\lambda=0\}})}{\frac{1}{n}-0}=lim_{t\rightarrow\infty}n\mu(T_t A_1\cap T_t A_{\frac{1}{2}} \cap...\cap T_t A_{\frac{1}{n}}))=n\mu( A_1)\mu(A_2)...\mu(A_{\frac{1}{n}})\leq n\mu(A_1)^n\longrightarrow0$ when $n\rightarrow\infty$, and where we have used that $\mu_{M}|_{A_{\lambda}}$ is mixing in the second equality. Therefore,  $lim_{n\rightarrow\infty}\frac{\mu (A_{\frac{1}{n}})-\mu(A_{\{\lambda=0\}})}{\frac{1}{n}-0}=0$, and because we assume that $\mu (A_{\lambda})$ is continuous for all $\lambda$, we conclude that $\frac{\partial \mu (A_{\lambda})}{\partial \lambda}|_{\lambda=0}=0$. Now if we expand $\mu (A_{\lambda})$ in power series around $\lambda=\lambda_c$ with the critical condition $\frac{\partial^{2} \mu (A_{\lambda})}{\partial \lambda^{2}}|_{\lambda=\lambda_c}=0$ we have $\mu (A_{\lambda})\simeq\mu (A_{\lambda_c})+\alpha_1(\lambda-\lambda_c)+\alpha_3(\lambda-\lambda_c)^3$ neglecting terms of order $\mathcal{O}(|\lambda-\lambda_c|^4)$. Now if we impose the conditions $\frac{\partial \mu (A_{\lambda})}{\partial \lambda}|_{\lambda=0}=\mu(A_{\{\lambda=0\}})=0$ we have $\alpha_1=\frac{3\mu (A_{\lambda_c})}{2\lambda_c}$ and $\alpha_3=-\frac{3\mu (A_{\lambda_c})}{\lambda_c^3}$. Therefore, if we replace these values in $\mu (A_{\lambda})$ we obtain $\mu(A_{\lambda})\simeq\mu(A_{\lambda_c})\left(\frac{3}{2}(\frac{\lambda}{\lambda_c})^{2}-\frac{1}{2}(\frac{\lambda}{\lambda_c})^{3}\right)$. Finally, from this formula we have that for small perturbations $\lambda\ll1$: $\mu(A_{\lambda})\simeq \mu(A_{\lambda_c})\frac{3}{2}(\frac{\lambda}{\lambda_c})^{2}$. On the other hand, we can expand $\frac{vol(\mathcal{C}_{\lambda}^{(M)})}{vol(\mathcal{C})}$ in power series of $\mu (A_{\lambda})$:
$\frac{vol(\mathcal{C}_{\lambda}^{(M)})}{vol(\mathcal{C})}=a_0+a_1\mu (A_{\lambda})+a_2\mu (A_{\lambda})^2+...=a_0+a_1\left(\mu(A_{\lambda_c})\left(\frac{3}{2}(\frac{\lambda}{\lambda_c})^{2}-\frac{1}{2}(\frac{\lambda}{\lambda_c})^{3}\right)\right)+...$ and since $\frac{vol(\mathcal{C}_{\lambda}^{(M)})}{vol(\mathcal{C})}(\lambda=0)=0$ (absence of chaos when $\lambda=0$) for small perturbations $\lambda\ll1$ we have $a_0=0$ and $\frac{vol(\mathcal{C}_{\lambda}^{(M)})}{vol(\mathcal{C})}\simeq a_1\mu(A_{\lambda_c})\frac{3}{2}(\frac{\lambda}{\lambda_c})^{2}\propto\mu(A_{\lambda})$.
\end{proof}
\end{theorem}

\begin{theorem}(Characterizing chaos with distances in the space state)\label{teo4}
If $N$ is the dimension of the Hilbert space and $d(\widehat{A},\widehat{B})=\left(Tr((\widehat{A}-\widehat{B})(\widehat{A}-\widehat{B})^{\dag})\right)^{\frac{1}{2}}$ is the distance between two operators $\widehat{A}$ and $\widehat{B}$, when $\hbar\sim0$ (classical limit) we have
\begin{equation}\label{teo4-1}
\left(d^{2}\left(\widehat{\rho}_{A_{\lambda}},\frac{1}{N}\textbf{1}\right)+\frac{1}{N}\right)\mu(A_{\lambda})=1
\end{equation}
\noindent where $\mu(A_{\lambda})=Tr(\widehat{I}_{A_{\lambda}})$, $\widehat{\rho}_{A_{\lambda}}=\frac{1}{\mu(A_{\lambda})}\widehat{I}_{A_{\lambda}}$, $\widehat{I}_{A_{\lambda}}=symb^{-1}(I_{A_{\lambda}})(\phi)$\footnote{Where $symb: \mathcal{A}\rightarrow \mathcal{A}_q$ is the Wigner transformation, $\mathcal{A}$ is the quantum algebra and $\mathcal{A}_q$ is a ``classical-like" algebra. See \cite{Wigner} for more details.} and $I_{A_{\lambda}}(\phi)$ is the characteristic function of $A_{\lambda}$. If $N=\infty$ we have $d^{2}(\widehat{\rho}_{A_{\lambda}},\frac{1}{N}\textbf{1})\mu(A_{\lambda})=1$.
\begin{proof}
If we make $\widehat{A}=\frac{1}{N}\textbf{1}$ and $\widehat{B}=\widehat{\rho}_{A_{\lambda}}$ we have
$d^{2}(\widehat{\rho}_{A_{\lambda}},\frac{1}{N}\textbf{1})=Tr((\widehat{\rho}_{A_{\lambda}}-\frac{1}{N}\textbf{1})(\widehat{\rho}_{A_{\lambda}}-\frac{1}{N}\textbf{1})^{\dag})=
Tr(\frac{1}{N^2}\textbf{1}-\frac{2}{N}\widehat{\rho}_{A_{\lambda}}+(\widehat{\rho}_{A_{\lambda}})^2)=\frac{1}{N^2}Tr(\textbf{1})-\frac{2}{N}Tr(\widehat{\rho}_{A_{\lambda}})+Tr((\widehat{\rho}_{A_{\lambda}})^2)$
=$\frac{1}{N^2}N-\frac{2}{N}+\frac{1}{\mu(A_{\lambda})^2}Tr((\widehat{I}_{A_{\lambda}})^2)=\frac{1}{N}-\frac{2}{N}+\frac{1}{\mu(A_{\lambda})^2}Tr(\widehat{I}_{A_{\lambda}})=-\frac{1}{N}+\frac{1}{\mu(A_{\lambda})^2}\mu(A_{\lambda})$
=$\frac{1}{\mu(A_{\lambda})}-\frac{1}{N}$ $\Longrightarrow d^{2}(\widehat{\rho}_{A_{\lambda}},\frac{1}{N}\textbf{1})=\frac{1}{\mu(A_{\lambda})}-\frac{1}{N}$ $\Longrightarrow \left(d^{2}(\widehat{\rho}_{A_{\lambda}},\frac{1}{N}\textbf{1})+\frac{1}{N}\right)\mu(A_{\lambda})=1$, where we have used that $(\widehat{I}_{A_{\lambda}})^2=\widehat{I}_{A_{\lambda}}$ when $\hbar\approx0$\footnote{When $\hbar\approx0$ we have $symb(\widehat{A}.\widehat{B})\cong symb(\widehat{A})symb(\widehat{B})$ for all $\widehat{A},\widehat{B}\in\mathcal{A}$.}. Finally, if $N=\infty$ we have that $d^{2}(\widehat{\rho}_{A_{\lambda}},\frac{1}{N}\textbf{1})\mu(A_{\lambda})=1$.

\end{proof}
\end{theorem}

\section{Physical relevance}

We illustrate the physical relevance of the framework with an example of the quantum chaos literature, the kicked rotator \cite{stockmann,haake,casati}, whose Hamiltonian $H=L^{2}+\lambda \ cos\theta\sum_n \delta(t-n)$ describes the free rotation of a pendulum with angular momentum $L$ periodically kicked by a gravitational potential of strength $\lambda$. The moment of inertia $I$ and the kick period $T$ are normalized to one and the evolution operator
is given by the Floquet operator $F=e^{\frac{-i}{\hbar}\lambda cos\theta}e^{\frac{-i}{2\hbar}\tau L^{2}}$ (see \cite{stockmann} pag. 147). Given an initial state $\rho(0)$ in the eigenbasis $\{|k\rangle\}$ of $F$ expressed by $\rho(0)=\sum_k \rho_{kk} |k\rangle\langle k|+\sum_{k\neq
k^{\prime}}\rho_{kk^{\prime}}|k\rangle\langle k^{\prime}|$, after $N$ successive applications of $F$ 
we have $\rho(N\tau)=F^{N}\rho(0)(F^{N})^{\dag}=\sum_k \rho_{kk} |k\rangle\langle k|+\sum_{k\neq
k^{\prime}}\rho_{kk^{\prime}}e^{-iN(\phi_k-\phi_{k^{\prime}})}|k\rangle\langle k^{\prime}|$
where the first and the second sums are
the diagonal and non-diagonal terms of the state $\rho(N\tau)$, and
the phase $e^{-iN\phi_k}$ is the eigenvalue of the eigenstate
$|k\rangle$ (see \cite{stockmann} pag. 137). Let $O$ be an
observable. We have that the mean value of
$O$ in the state $\rho(N\tau)$ is $\langle O\rangle_{\rho(N\tau)}=(\rho(N\tau)|O)=\sum_k
\rho_{kk} O_{kk}+\sum_{k\neq k^{\prime}}\rho_{kk^{\prime}}e^{-iN(\phi_k-\phi_{k^{\prime}})}O_{kk^{\prime}}$ where the second sum contains the purely quantum correlations. 
Now if we use that $lim_{N\rightarrow\infty}\frac{1-e^{-iN(\phi_k-\phi_{k^{\prime}})}}{N(1-e^{-i(\phi_k-\phi_{k^{\prime}})})}=0$
for all $k\neq k^{\prime}$ and that $\rho_{\ast}=\sum_k
\rho_{kk} |k\rangle\langle k|$ we have that $lim_{N\rightarrow\infty}\frac{1}{N}\sum_{j=0}^{N-1}(\rho(j\tau)|O)$ is equals to $\sum_k\rho_{kk} O_{kk}=(\rho_{\ast}|O)$. From this and the definition of the ergodic level (Eqns. \eqref{qcorrelation} and \eqref{qcorrelation2}) we have that for all $\lambda$ the kicked rotator is
ergodic. The C\`{e}saro limit is $\rho_{\ast}=\sum_k \rho_{kk}
|k\rangle\langle k|$ which is the equilibrium state ``on
time-average" 
. In terms of the convex subsets we have that
$\mathcal{C}_{\lambda}^{(E)}=\mathcal{C}$ for all $\lambda$. 

On the other hand, for $\lambda>5$ the behavior is
different than in the ergodic case. In this case the quadratic mean value of
the angular momentum $\langle L^{2}\rangle$ after $N$ kicks is $f_N(L)=\frac{1}{l_s}e^{-\frac{2|L|}{l_s}}$ (see \cite{stockmann} pag. 149). This exponential localization implies that for kick numbers $N\leq l_s$ we are in the range of classical diffusion and
for $N\gg l_s$ we are in the fully chaotic behavior. For $N\gg l_s$
the phase factors $e^{-iN(\phi_k-\phi_{k^{\prime}})}$ rapidly oscillate that only the terms with
$k=k^{\prime}$ survive, that is $\langle O\rangle_{\rho(N\tau)}=(\rho(N\tau)|O)\simeq \sum_k
\rho_{kk} O_{kk}=(\rho_{\ast}|O)$ where $\rho_{\ast}=\sum_k \rho_{kk} |k\rangle\langle k|$.
Then when $\lambda>5$ we have $lim_{N\rightarrow\infty}(\rho(N)|O)=lim_{N\rightarrow\infty}(\rho(N\tau)|O)=\sum_k\rho_{kk} O_{kk}=(\rho_{\ast}|O)$. This means that for $\lambda>5$ the kicked rotator belongs to
the mixing level (Eqns.  \eqref{qcorrelation} and \eqref{qcorrelation2}) and the equilibrium state is the weak limit
$\rho_{\ast}=\sum_k \rho_{kk} |k\rangle\langle k|$ which is
the diagonal part of the initial state $\rho(0)$ written in the basis of the Floquet operator $F$ 
. Meanwhile, for the fully chaotic regime $\lambda>5$ the kicked rotator relaxes in the Floquet basis
$\{|k\rangle\}$. In terms of the convex subsets we have that
$\mathcal{C}_{\lambda}^{(M)}=\mathcal{C}$ for all $\lambda>5$.

\section{Conclusions}
We have presented a geometrical approach to mixed characterized by a Hamiltonian $H=H_0+\lambda V$ with $H_0$ and $\lambda V$ the unperturbed and perturbed parts and provided with a phase space decomposed into two uncorrelated regions $A_{\lambda}$ (chaotic) and $E_{\lambda}$ (integrable). Our proposal has the advantage of connecting the intuitive notion of a mixed space with the mixing level of the EH.
Using the EH and QEH levels we have shown some features of the mixed systems: 1) the chaos transitions in terms of the deformations of $\mathcal{C}^{(E,M)}$ , 2) the quadratic increase of chaos for small perturbations and 3) a relationship between the measure $\mu(A_{\lambda})$ of the chaotic region $A_{\lambda}$ and the distance between the maximally mixed state $\frac{1}{N}\textbf{1}$ and $\rho_{A_\lambda}$ (the associated state of $A_{\lambda}$).
Our geometrical approach offers three possible ``measures" of the chaos transitions of the mixed systems: $\mu(A_{\lambda})$, $\frac{vol(\mathcal{C}_{\lambda}^{(M)})}{vol(\mathcal{C})}$ and $d^{2}(\widehat{\rho}_{A_{\lambda}},\frac{1}{N}\textbf{1})$. However, the model can not predict the constants $\mu(A_{\lambda_c})$ and $\lambda_c$ because these depend on the system.
For the kicked rotator case the critical value is $\lambda_c=0,9716...$
so employing 
\eqref{teo1-2}
we obtain $\frac{\mu(A_{\{\lambda=0.2\}})}{\mu(A_{\lambda_c})}\approx 0,05919$ which means that when $\lambda=0.2$ the chaotic region $A_{\lambda}$ occupies approximately a sixth percent of the total volume of the phase space. This is in a reasonable agreement with numerical calculations (see \cite{stockmann} Fig. 1.3 (a)) and can be considered as a first test of the presented framework.

\section*{Acknowledgments}
Ignacio S. Gomez acknowledges support from the Department of Exact
and Natural Sciences of the State University of Southwest
Bahia (UESB), Itapetinga, Bahia, Brazil and from the Conselho Nacional de Desenvolvimento Científico e Tecnológico (CNPq), Grant Number
316131/2023-7. 



\begin{thebibliography}{0}

\bibitem{stockmann} H. Stockmann, \textit{Quantum Chaos - An Introduction}, Cambridge Univ. Press, Cambridge, 1999.

\bibitem{haake} F. Haake, \textit{Quantum Signatures of Chaos,} 2nd edition, Springer-Verlag, Heidelberg, 2001.

\bibitem{gutzwiller} M. C. Gutzwiller, \textit{Chaos in Classical and Quantum Mechanics}, Springer Verlag, New York, 1990.

\bibitem{casati} G. Casati, B. Chirikov, \textit{Quantum Chaos: between order and disorder,} Cambridge Univ. Press, Cambridge 1995.

\bibitem{casati verdadero} G. Casati, T. Prosen, \textit{Phys. Lett. A}, 72, 032111, 2005.

\bibitem{casati model} M. Castagnino, \textit{Phys. Lett. A}, 357, 97-100, 2006.

\bibitem{tabor} M. Tabor, \textit{Chaos and Integrability in Nonlinear Dynamics: An Introduction}, Wiley, New York, 1988.

\bibitem{BELOT} G. Belot, J. Earman, \emph{Stud. His.
Philos. Mod. Phys.}, \textbf{28}, 147-182, 1997.

\bibitem{lich} A. J. Lichtenberg, M. A. Lieberman, \textit{Regular and Chaotic Dynamics}, Springer Verlag, New York, 1992.


\bibitem{3M} J. Berkovitz, R. Frigg, F. Kronz, \textit{Stud. Hist. Phil.
Mod. Phys.,} \textbf{37}, 661-691, 2006.

\bibitem{0} M. Castagnino, O. Lombardi, \textit{Physica A}, \textbf{388},
247-267, 2009.

\bibitem{NACHOSKY MARIO} I. Gomez,  M. Castagnino, \textit{Physica A}, \textbf{393},
112--131, 2014.

\bibitem{vkampen} N. van Kampen, \textit{Phys. A}, \textbf{20}, 603, 1969.

\bibitem{daneri} A. Daneri, A. Loinger, G. Prosperi, \textit{Nucl. Phys.}, \textbf{33}, 267, 1962.

\bibitem{zeh} H. Zeh, \textit{Found. Phys.}, \textbf{3}, 109, 1973.

\bibitem{vhove}

L. van Hove, \textit{Phys. A}, \textbf{20}, 603, 1954.

L. van Hove, \textit{Phys. A}, \textbf{25}, 268, 1959.


\bibitem{movhove} M. Castagnino, O. Lombardi, \textit{Stud. Hist. Phil. Mod. Phys.}, \textbf{35}, 73, 2004.

\bibitem{MielnikGQS} B. Mielnik, \textit{Commun. math. Phys.}, \textbf{15}, 1-46, 1969.

\bibitem{Holik} F. Holik, M. Plastino, \emph{Phys. Rev. A}, \textbf{84}, 062327

\bibitem{Omnes} R. Omn\`{e}s, \textit{The Interpretation of Quantum Mechanics}, Princeton University, Princeton, 1994.

\bibitem{nacho1}
I. Gomez, M. Castagnino, \textit{Chaos, Solitons \& Fractals}, \textbf{68}, 98--113 2014.

\bibitem{nacho2} I. Gomez, M. Castagnino, \textit{Chaos, Solitons \& Fractals}, \textbf{70}, 99--116 2015.

\bibitem{nacho3} I. S. Gomez, M. Portesi, \textit{Physica A}, \textbf{479}, 437--448, 2017.

\bibitem{nacho4} I. S. Gomez, \textit{Chaos, Solitons \& Fractals}, \textbf{106}, 317--322, 2018.


\bibitem{nacho5} I. S. Gomez, \textit{Int. J. Bif. Chaos}, \textbf{28}, 1850002, 2018.

\bibitem{nacho6} I. S. Gomez, E. P. Borges, \textit{J. Stat. Mech.}, 063105, 2018.

\bibitem{nacho7} I. Gomez, M. Losada, S. Fortin, M. Castagnino, M. Portesi, \emph{Int. J. Theor. Phys.} \textbf{54}, 2192--2203, 2015.

\bibitem{nacho8}  I. S. Gomez, \textit{Chaos, Solitons \& Fractals}, \textbf{99}, 155--171, 2017.

\bibitem{Bro73} T. Brody, A statistical measure for the repulsion of energy levels, \textit{Lettere Al Nuovo Cimento}, \textbf{7}, 482, 1973.

\bibitem{Len91} G. Lenz, F. Haake, \textit{Phys. Rev. Lett.}, \textbf{67}, 1, 1991.

\bibitem{Rob83} M. Robnik, Classical dynamics of a family of billiards with analytic boundaries, \textit{J. Phys. A}, \textbf{16}, 3971, 1983.

\bibitem{Ber84} M. Berry, M. Robnik, \textit{J. Phys. A}, \textbf{17}, 2413, 1984.


\bibitem{bruselas} N. Bleistein, R. Handelsman, \textit{Asymptotic Expansion of Integrals}, Dover, New York, 1986.

\bibitem{mario chaos} M. Castagnino, O. Lombardi, \textit{Chaos, Solitons and Fractals}, \textbf{28}, 845-1112, 2006.

\bibitem{Wigner} M. Hillery, R. O'Connell, M. Scully, E. Wigner, \textit{Phys.
Rep}. \textbf{106}, 121-167 (1984).





\end{thebibliography}
\end{document}